\documentclass[usenatbib]{mn2e}

\usepackage{graphicx}
\usepackage{aastex_hack}
\usepackage{color}

\usepackage[hyphens]{url}
\usepackage{hyperref}

\def\eg{e.g., }

\newcommand{\beq}{\begin{equation}}
\newcommand{\eeq}{\end{equation}}

\def\hyi{\ion{H}{1}}

\def\oiilam{[\ion{O}{2}]~$\lambda3727$}

\def\oiiilam{[\ion{O}{3}]~$\lambda5007$}

\def\mgii{\ion{Mg}{2}}

\def\nai{\ion{Na}{1}}
\def\naii{\ion{Na}{2}}
\def\cai{\ion{Ca}{1}}
\def\caii{\ion{Ca}{2}}
\def\caiii{\ion{Ca}{3}}

\def\caiidoublet{\ion{Ca}{2}~$\lambda\lambda3934,3969$}

\def\naidoublet{\ion{Na}{1}~$\lambda\lambda5890,5896$}

\newcommand{\kms}{\ensuremath{{\rm km~s}^{-1}}}

\newcommand{\rewcaiitwo}{\ensuremath{W_0^{\rm H}}}

\newcommand{\rewnaitwo}{\ensuremath{W_0^{\rm D_1}}}

\newcommand{\caplus}{\ensuremath{\rm{Ca^+}}}

\newcommand{\vlsr}{\ensuremath{v_{\rm LSR}}}
\newcommand{\ebv}{\ensuremath{E_{B-V}}}
\newcommand{\Nhyi}{\ensuremath{N_{\rm{H\,I}}}}
\newcommand{\Nnai}{\ensuremath{N_{\rm{Na\,I}}}}
\newcommand{\Ncaii}{\ensuremath{N_{\rm{Ca\,II}}}}

\title[Ca\,II and Na\,I absorption by the Milky Way]{Calcium H \& K and sodium D absorption induced by the interstellar and circumgalactic media of the Milky Way}

\author[Murga et al.]
{
\parbox[h]{\textwidth}{
Maria Murga$^{1}$,
Guangtun Zhu$^{2,\dagger}$,
Brice M{\'e}nard$^{2,3}$ \&
Ting-Wen Lan$^{2}$
}
\vspace*{2pt} \\
\hspace{-.1cm}$^1$ Institute of Astronomy of Russian Academy of Sciences, Pyatnitskaya str. 48, Moscow 119017, Russia; email: murga@inasan.ru\\
\hspace{-.1cm}$^2$ Department of Physics \& Astronomy, Johns Hopkins University, 3400 N. Charles Street, Baltimore, MD 21218, USA\\
\hspace{-.1cm}$^\dagger$ Hubble fellow\\
\hspace{-.1cm}$^3$ Kavli IPMU (WPI), the University of Tokyo, Kashiwa 277-8583, Japan \\
}

\begin{document}
\newcounter{tabcount}
\date{Draft, \today}

\maketitle

\begin{abstract}
We map out calcium II \& sodium I absorption (Fraunhofer H, K \& D lines) induced by both the interstellar medium and the circumgalactic medium of the Milky Way. Our measurements cover more than $9000$ deg$^2$ and make use of about $300,000$ extragalactic spectra from the Sloan Digital Sky Survey. We present absorption maps for these two species and then compare their distributions to those of neutral hydrogen and dust. We show that the abundance of \nai\ with respect to neutral hydrogen stays roughly constant in different environments, while that of \caii\ decreases with hydrogen column density. Studying how these tracers vary as a function of velocity, we show that, on average, the \Nnai/\Ncaii\ ratio decreases at higher velocity with respect to the local standard of rest, similar to the local Routly-Spitzer effect but seen on Galactic scale. We show that it is likely caused by higher gas/dust density at lower velocity. Finally, we show that Galactic \caii\ and \nai\ absorption needs to be taken into account for precision photometry and, more importantly, for photometric redshift estimation with star forming galaxies. Our maps of \caii\ and \nai\ absorption are publicly available.
\end{abstract}


\section{Introduction}\label{sec:intro}

Various advances in our understanding of the Universe have been enabled by sky surveys mapping emission and absorption features across the sky. Historically, William Herschel pioneered this approach by mapping the density of stars and revealing the existence of "dark holes" (Herschel, 1785), known today as dust clouds. The mapping of globular clusters across the sky allowed Harlow Shapley to infer the disk-like geometry of the Milky Way about one hundred years ago. In the sixties, radio surveys opened up a new window giving us access to the distribution of hydrogen in the Galaxy \citep[\eg][]{kalberla09a}. Over the past twenty years, IR observations have provided us with the distribution of dust \citep[\eg][]{schlegel98a} and CO observations that of molecular gas \citep[\eg][]{2001ApJ...547..792D}. More recently, over the past few years, large-scale maps of diffuse interstellar bands tracing (still unidentified) large molecules have started to emerge \citep{2014arXiv1406.7284L,2015ApJ...798...35Z,2015MNRAS.447..545B}.
Surprisingly, the large-scale distribution of metals in the atomic gas phase -- a fundamental component of the ISM -- remains poorly measured. Characterizing their spatial distribution is important: while metals are expected to represent about 2\% of the mass of the interstellar medium (ISM), they often determine the chemistry, ionization state and temperature of the gas. Their spatial distribution reflects the return to the ISM of gas that has been processed in stars and stellar explosions. As of today, no large-scale map of metals in the gas phase exists.

Gaseous metals can be probed using absorption line spectroscopy. Among the features accessible in the optical window, the \caii\ and \nai\ lines are the most prominent. They correspond to the H, K and D lines discovered by Fraunhofer in the solar spectrum two hundred years ago (Fraunhofer 1814). Over the last two centuries, these Fraunhofer lines have played an important role in the understanding of a variety of astronomical phenomena both in the Milky Way \citep[\eg][]{hartmann04a, routly52a} and on extragalactic scales \citep[\eg][]{boksenberg78a, blades81a}. In this paper we use a large set of extragalactic spectra observed by the Sloan Digital Sky Survey \citep[SDSS,][]{york00a} to measure the absorption induced by the Milky Way. Interestingly, while such information is encoded in virtually any extragalactic spectrum, no systematic extraction and analysis has been attempted on a large scale. Only \citet{poznanski12a} measured \nai\ absorption in SDSS galaxy spectra to characterize its cross-correlation with dust reddening. 
We create maps
\footnote{The maps of Ca$^{+}$ and Na absorption are made publicly available at:~\tt{\url{http://www.inasan.ru/~khramtsova/project_CaNa}}} 
of \caii\ and \nai\ absorption over a quarter of the sky induced by gas in and around the Milky Way, i.e., in both the ISM and the circumgalactic medium (CGM), and study their global correlations with other baryonic tracers, such as neutral hydrogen and dust grains.
%
The Ca H \& K doublet corresponds to the fine structure splitting of the singly ionized calcium excited states \caii\ (\caplus) at $\lambda=3934.78\,$\AA\ (K) \& $3969.59\,$\AA\ (H) in vacuum, and Na D, which is also a doublet, corresponds to the fine structure splitting of the neutral sodium excited states at $\lambda=5891.58\,$\AA\ (D$_2$) \& $5897.56\,$\AA\ (D$_1$) in vacuum. Since their discovery they have played important roles in astrophysics because of their strength and their location in the visible part of the spectrum. Both calcium and sodium are among the most abundant heavy elements, with $\log(\rm{Ca/H})_\odot+12 \simeq 6.34$ and $\log(\rm{Na/H})_\odot+12 \simeq 6.24$ (Asplund 2009), and are found in a variety of astrophysical environments. The ionization potentials are $6.11\,$eV, $11.87\,$eV and $50.91\,$eV for \cai, \caii\ and \caiii, and $5.14\,$eV and $47.29\,$eV for \nai\ and \naii\ \citep[][]{morton03a}. With ionization potentials lower than $1$ Rydberg, \caii\ and \nai\ are therefore not the dominant stage of ionization in most astrophysical environments.

The paper is organized as follows. In Section \ref{sec:data}, we briefly describe the data set and method. In Section \ref{sec:analysis}, we present the maps of \caii\ and \nai\ absorption and investigate the correlations of metal absorption with neutral hydrogen and dust. We discuss the effect of metal absorption on precision photometry in Section~\ref{sec:photometry}. Section~\ref{sec:summary} summarizes our main results.

\begin{figure*}
    \includegraphics[width=0.45\textwidth]{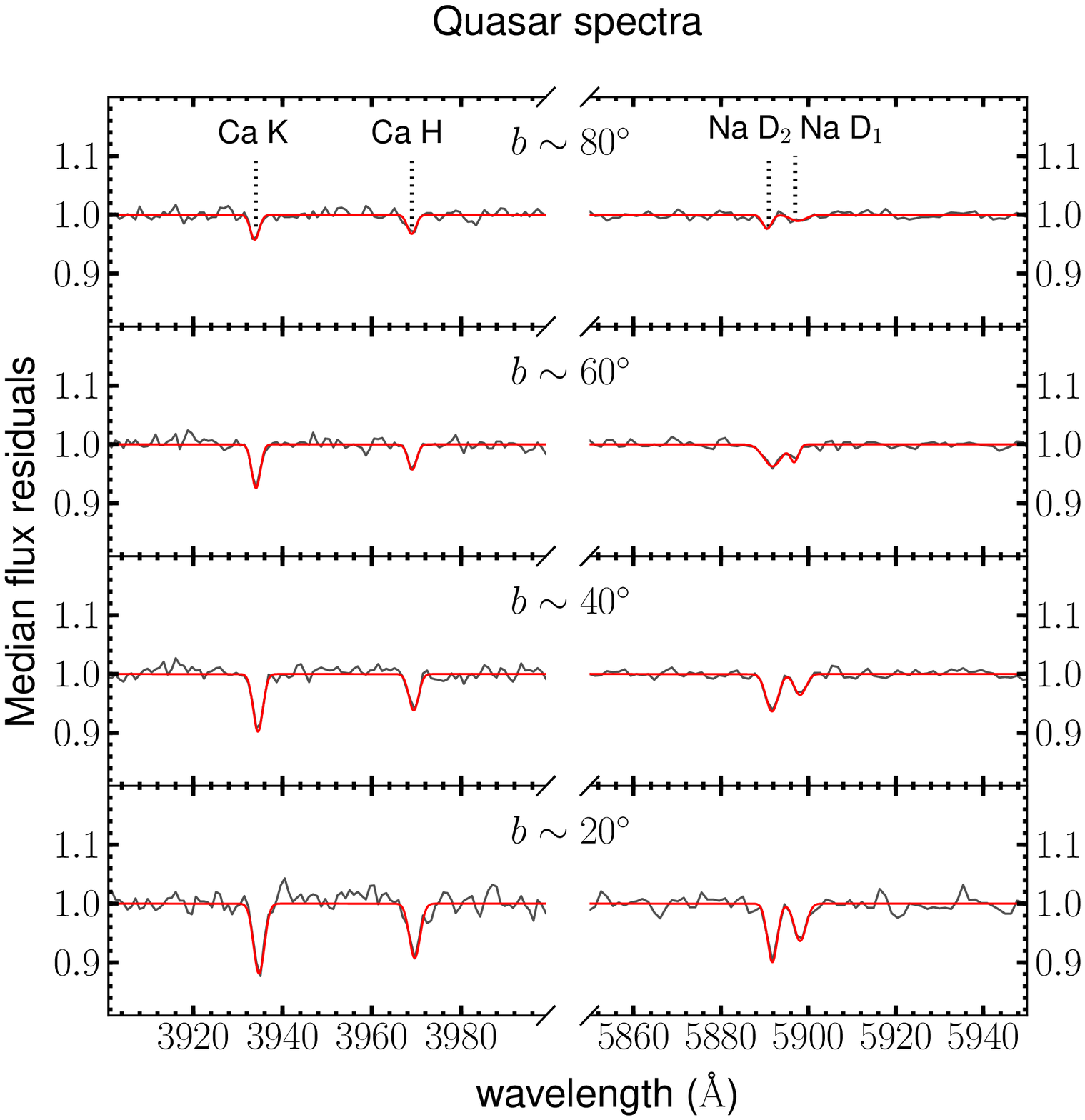}
    \includegraphics[width=0.45\textwidth]{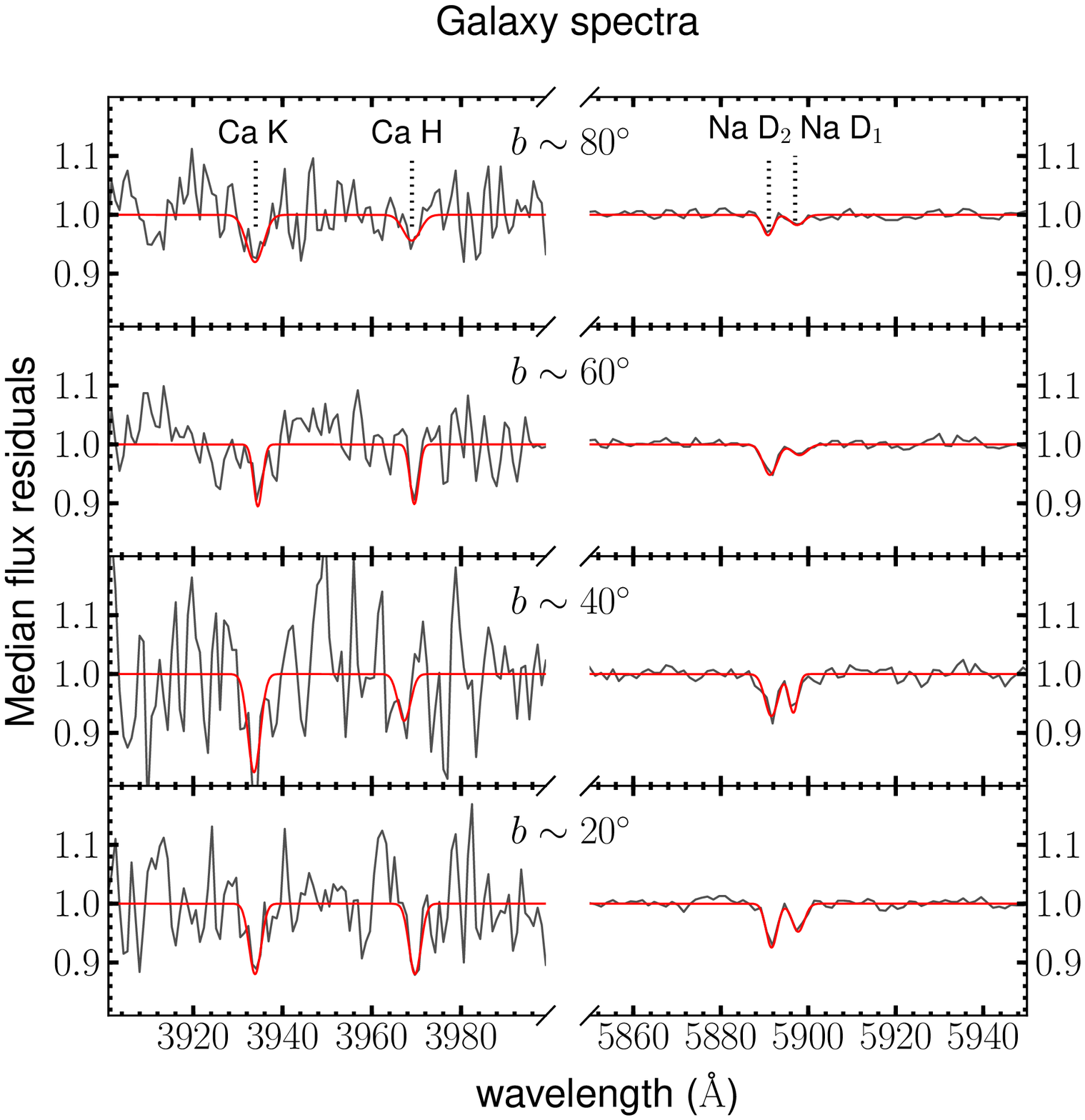}
    \caption{Examples of stacked continuum-normalized residual spectra of quasars (left panel) and galaxies (right panel) in four Galactic latitude bins.
The red lines represent the best-fit double-Gaussian profiles.}
\label{fig:stackedspectra}
\end{figure*}

\section{The data}\label{sec:data}

\begin{figure*}
\includegraphics[width=.8\textwidth]{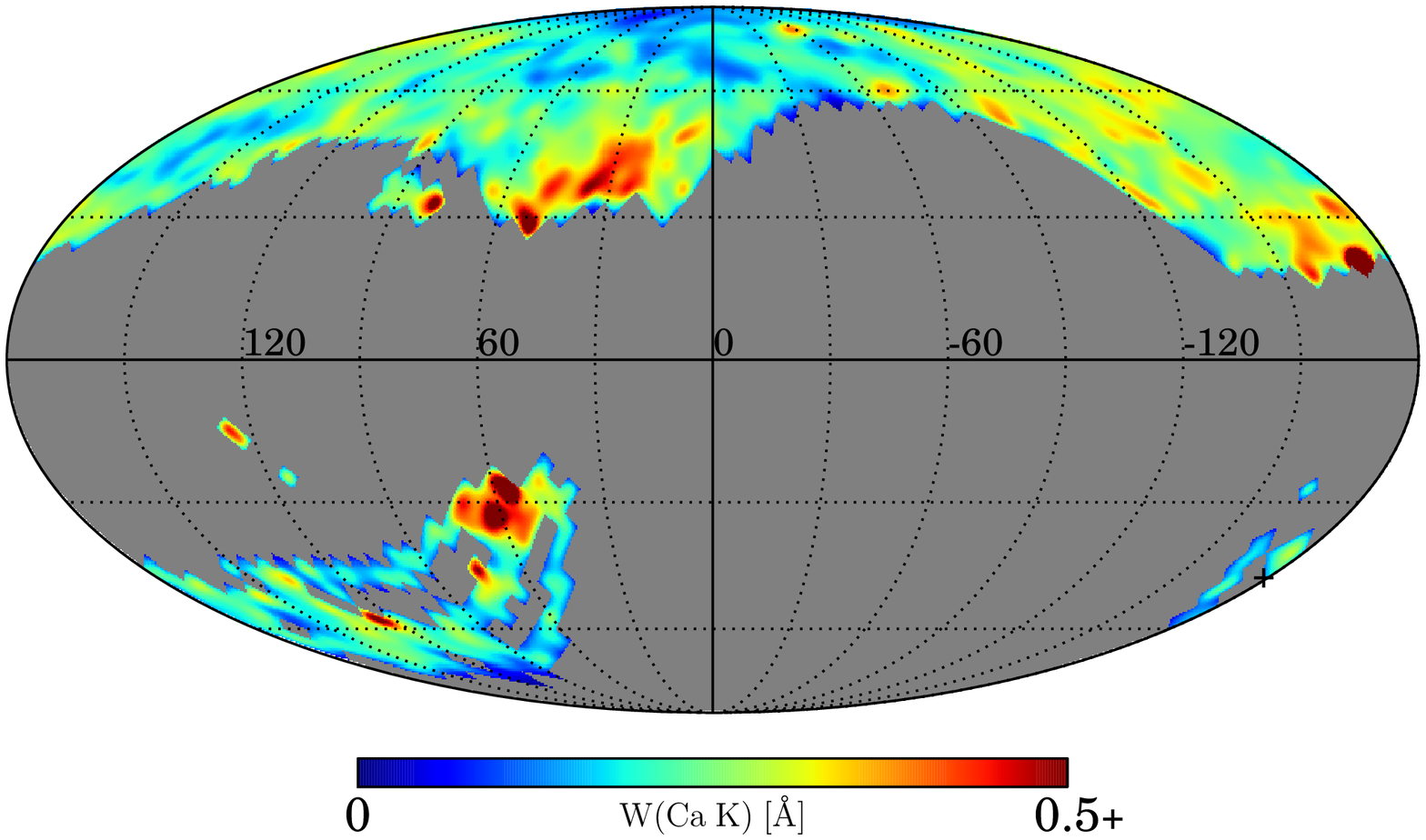}
\includegraphics[width=.8\textwidth]{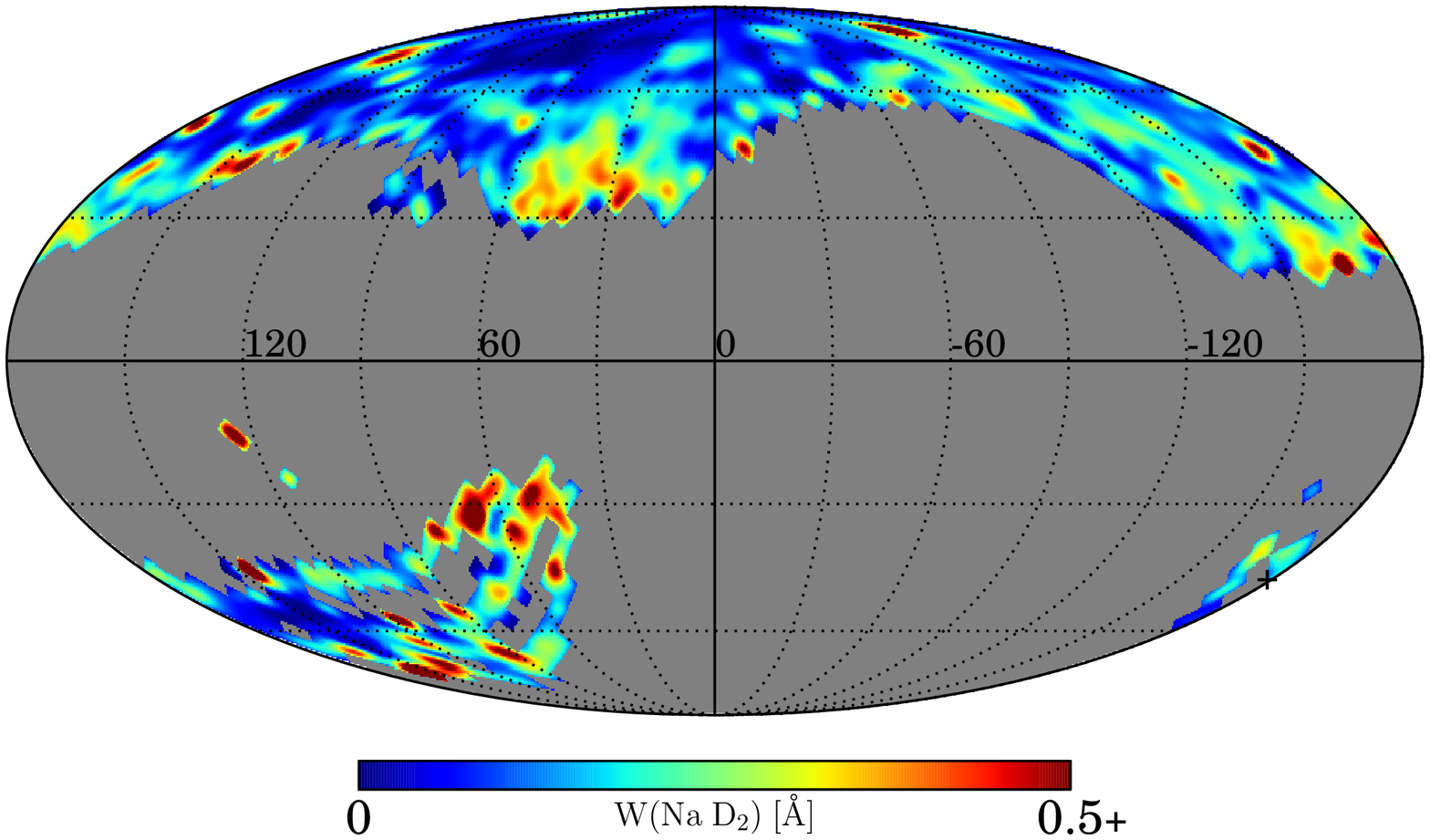}
\caption{Maps of \caii\ (top) and \nai\ (bottom) created from measurements of Fraunhofer H and ${\rm D}_2$ absorption lines in the spectra of about 300,000 quasars and galaxies from the Sloan Digital Sky Survey. These maps use the Mollweide projection and show the equivalent width of each line from 0 to $0.5\,$\AA\ using a 3-degree Gaussian smoothing. Note that to visually enhance the constrast, equivalent width values greater than $0.5\,$\AA\ are saturated. This affects only a small fraction of the pixels.}
\label{fig:maps}
\end{figure*}

We derive our absorption measurements from quasar and galaxy spectra taken from the SDSS Data Release 7 \citep[DR7,][]{abazajian09a}. For quasars, we use the quasar catalog compiled by \cite{schneider10a} and the improved redshift estimates given by \citet{hewett10a}. For galaxies, we use the MPA-JHU value-added catalog \citep[\eg][]{tremonti04a}. Absorption line measurements require accurate estimation of the intrinsic background source continuum $\hat{F} (\lambda)$. Below we briefly describe the continuum estimates we use:
\begin{itemize}
\item \textbf{Quasar continuum estimation}: We use the estimates provided by \citet{zhu13a}, who used a basis set of quasar eigenspectra created with the dimensionality-reduction technique {\it nonnegative matrix factorization} \citep[NMF,][]{lee99a, blanton07a}. Large-scale residuals not accounted for by the NMF basis set are removed with appropriate median filters. This set of flux residuals has been used to create a sample of about 50,000 absorber systems \citep{zhu13a}, to measure the total amount of \caii\ around low-redshift galaxies \citep{zhu13b} and \mgii\ around luminous red galaxies at $z\sim0.5$ \citep{zhu14a}.
\item \textbf{Galaxy continuum estimation}: We use estimates of their intrinsic continumn fluxes provided by \citet{zhu10a}. These authors modeled the observed galaxy spectra using single stellar population (SSP) models of \citet{bruzual03a} with the Padova 1994 library of stellar evolution tracks \citep[\eg][]{alongi93a} and the \citet{chabrier03a} initial mass function (IMF). For this analysis, we restrict the selection of galaxies to those at redshift $>0.2$, most of which are so-called luminous red galaxies (LRGs) with well-understood spectral energy distribution. Because red galaxies have low-level fluxes blueward of the $4000\,$\AA\ break, there is limited information of \caii\ absorption induced by the foreground sources, we will use galaxy spectra only for \nai\ absorption.
\end{itemize}
In total, we make use of spectra of $84,533$ quasars and $181,286$ galaxies. These sources cover about a quarter of the sky, mostly at high Galactic latitude in the northern hemisphere with relatively low dust extinction. We note that all the background sources used here are extragalactic objects and the spectral resolution is $69\,\kms$.
All the absorption measurements presented here therefore include contributions from all gas, including both in the ISM and the CGM, of the Milky Way.\\

To cross-correlate our maps of metal absorption with the distributions of hydrogen and dust we use the following datasets:
\begin{itemize}
\item \textbf{Neutral hydrogen}: The Leiden/Argentine/Bonn (LAB) Galactic \hyi\ Survey \citet{kalberla09a} merged the Leiden/Dwingeloo Survey \citep[LDS:][]{hartmann97a} in the northern sky and the Instituto Argentino de Radioastronom\'ia Survey \citep[IAR:][]{arnal00a, bajaja05a} in the south and present an all-sky brightness temperature map of the $21\,{\rm cm}$ hyperfine emission by neutral hydrogen. The LAB Survey is the most sensitive Milky Way \hyi\ survey to date, with the most extensive coverage both spatially and kinematically. In the optically thin regime, the integrated emission yields total neutral hydrogen column density \Nhyi\ in the Milky Way, which we use to compare with our statistical metal absorption measurements. For \Nhyi, we use the values integrated over the full velocity range, from $-450\,\kms$ to $400\,\kms$.
\item \textbf{Dust}: \citet{schlegel98a} derived Galactic reddening \ebv\ over the full sky from a composite $100\,\mu{\rm m}$ of the COBE/DIRBE \citep[\eg][]{boggess92a} and IRAS/ISSA maps \citep[\eg][]{wheelock94a}, assuming uniform dust properties with values of $R_V=A_V/\ebv\approx3.1$. We here use \ebv\ as a proxy of line-of-sight dust column density and adopt their derivations. 
\end{itemize}
For both \Nhyi\ and \ebv, we use the maps distributed on the NASA LAMBDA website\footnote{\tt\url{http://lambda.gsfc.nasa.gov/product/foreground}}, which were produced under the HEALPix scheme \citep{gorski05a}. The HEALPix pixelization divides a spherical surface into pixels with the same area. The main parameter of the pixelization is the number of pixels along a fixed longitude $N_{\rm side}$. Both \Nhyi\footnote{These data were used in \citet{land07a} and given for public use through LAMBDA.} and \ebv\ maps were stored with $N_{\rm side}=512$, which divides the whole sky into 3,145,728 pixels, with the resolution of approximately $0.01\,$sq. deg per pixel. 

\section{Analysis}
\label{sec:analysis}

Given the typical signal-to-noise ratio of SDSS quasar and galaxy spectra, the \caii\ and \nai\ absorption lines induced by the Milky Way are in general not detectable in individual spectra. Their detection requires constructing averaged continuum-normalized residual spectra. To do so we use an inverse-variance weighted estimator, using the wavelength-dependent errors given by the SDSS pipeline. With the stacked residual spectra we then measure the rest equivalent width of the \caii\ and \nai\ doublets with a double-Gaussian profile, allowing the width and line ratio to be free parameters. In Figure~\ref{fig:stackedspectra}, we present examples of stacked spectra as a function of Galactic latitude, showing clear detections of \caii\ and \nai\ absorption. The best-fit double-Gaussian profiles are presented with the red lines. As shown in the right panel of the Figure, there is limited information about \caii\ absorption using galaxy spectra. This is because we are using red galaxies which have little fluxes blueward of the $4000\,$\AA\ break. In the following, we therefore use only quasars as background sources for \caii, while for \nai\, we combine both quasars and galaxies as a whole sample.

\subsection{Absorption Maps}

\begin{figure}
    \includegraphics[width=0.45\textwidth]{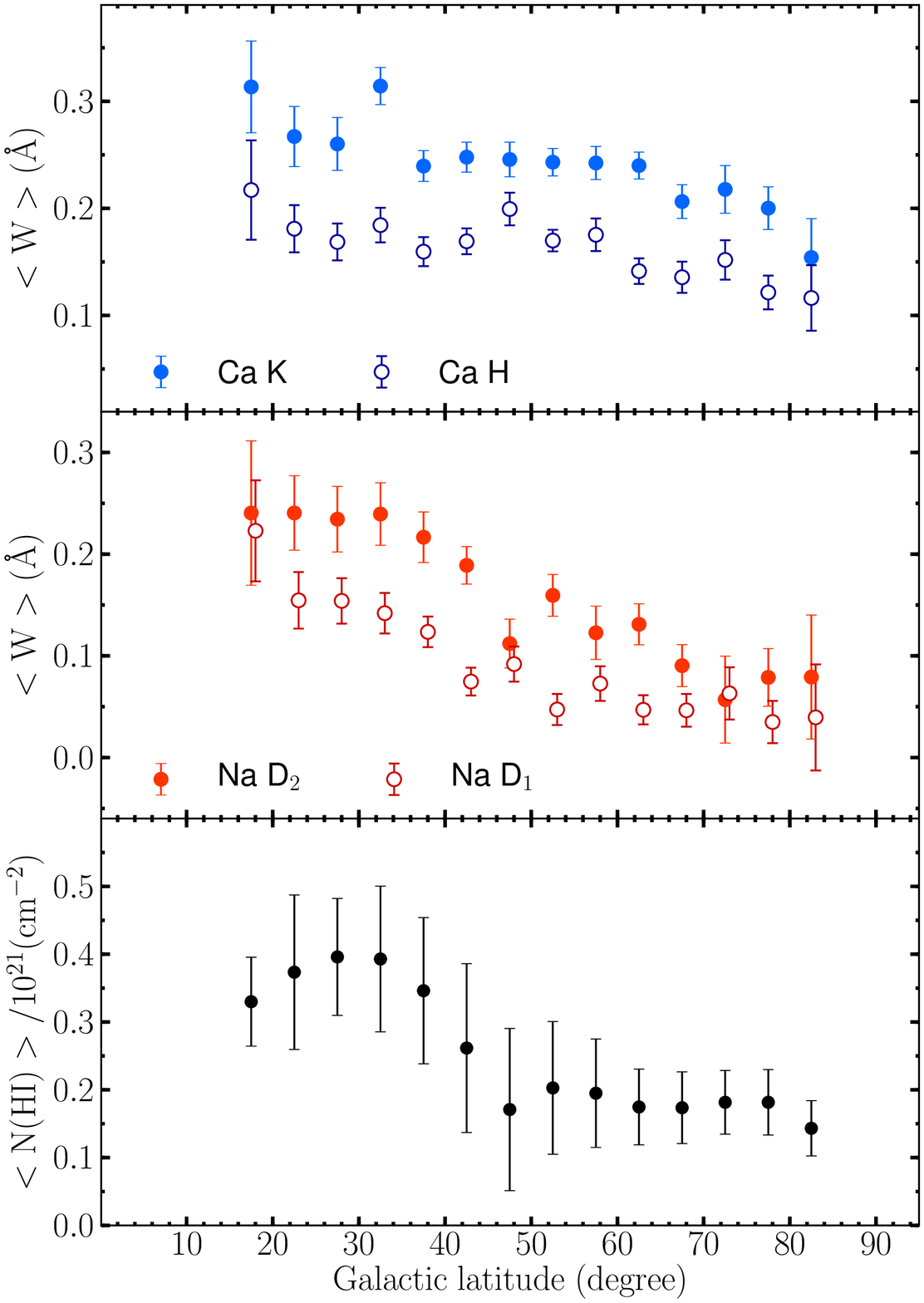}
    \caption{Rest equivalent width of \caii\ K (top panel) and \nai\ D$_2$ (middle panel) as a function of Galactic latitude $b$. The errorbars represent $1\sigma$ errors from the double-Gaussian fitting. For comparison, in the bottom panel we show the average neutral hydrogen column density \Nhyi\ in the same footprint, using values from the LAB \hyi\ survey. The errorbars for \Nhyi\ are $1\sigma$ dispersion in the pixels used.}
\label{fig:ewb}
\end{figure}

To create maps of \caii\ and \nai\ absorption we pixelize the sky using the HEALPix scheme. We choose the number of pixels along the Galactic longitude to be $N_{\rm side}=16$, which divides the whole sphere into $3072$ pixels with equal area of about $3.7^\circ\times3.7^\circ$. This choice of resolution is motivated by the surface number density of objects so that in the majority of the pixels, we have enough objects for a signal-dominated determination of the absorption strength. In total, our data cover about $700$ pixels (about $9,100\,{\rm deg}^2$), and we have on average about $120$ quasars and $270$ galaxies in a pixel. For each of these resolution elements we perform a double-Gaussian fitting to the stacked residual spectra and measure the rest equivalent width of each line. At that resolution, the typical S/N with which such lines are detected is about $3$.

Figure~\ref{fig:maps} presents maps of \caii\ K and \nai\ D$_2$ absorption in the Mollweide projection. They show the equivalent width of each line from 0 to $0.5\,$\AA\ using a 3-degree Gaussian smoothing. Note that to visually enhance the constrast, equivalent width values greater than $0.5\,$\AA\ are saturated. This affects only a small fraction of the pixels. Both maps show large-scale gradients as a function of Galactic latitude. The distribution of \nai\ appears to be more concentrated towards the disk of the Milky Way. 
To characterize this dependence, we measure the mean absorption strength as a function of latitude, using only data from the Northen Galactic hemisphere where the coverage is more complete. We present the mean equivalent width of \caii\ and \nai\ measured in bins with $\Delta b=10^\circ$ in Figure~\ref{fig:ewb}, where the errorbars show the $1\sigma$ dispersion of all the pixels. For comparison, we also show the average \hyi\ column density in the bottom panel, where \Nhyi\ is the median value over all pixels of the \Nhyi\ HEALPix map in each latitude bin within the same SDSS footprint. Both the \caii\ and \nai\ absorption, and \Nhyi\ decrease from low latitude towards high latitude. This is likely due to a combination of longer path through the disk at low latitude and higher local column density in the ISM.

\subsection{Correlations with Neutral Hydrogen and Dust}

More information on the distributions of \caii\ and \nai\ can be extracted from these maps using cross-correlations with other tracers. This allows one to make use of signal lying below the noise level of individual HEALPix pixels. In this section we present measurements of cross-correlations with the observed distributions of Galactic neutral hydrogen and dust. We note here that previous attempts to measure correlations between metal absorption, hydrogen and dust were typically based on datasets targetting ``dense" clouds. By using spectra of extragalactic sources and integrating the absorption signal over a large number of lines-of-sight, our approach allows us to probe the total amount of absorption originating from both dense clouds as well as the diffuse interstellar and circumgalactic media. So far, only one such approach has been used: \citet{poznanski12a} performed a similar statistical study of \caii\ absorption.

To investigate the dependence of \caii\ and \nai\ absorption on \Nhyi\ and \ebv, we use the high-resolution HEALPix maps of \Nhyi\ and \ebv\ distributed by the LAMBDA website. We divide all the HEALPix pixels into bins with $\Delta \log_{10} \Nhyi=0.05$ and $\Delta \log_{10} \ebv = 0.05$. We then find all objects (quasars and galaxies for \nai, quasars for \caii) in the corresponding pixels, construct stacked continuum-normalized residual spectra and measure the equivalent width of \caii\ and \nai. Note that we do not spatially downgrade the \Nhyi\ and \ebv\ HEALPix maps (with $N_{\rm side}=512$) but instead we select objects in pixels with the same \Nhyi\ or \ebv\ values. In Figure~\ref{fig:CaNa_nh}, we present the correlations of the rest equivalent width of \caii\ and \nai\ as a function of \Nhyi\ and \ebv, where the errorbars are from the double-Gaussian fitting. Both \caii\ and \nai\ absorption strengths correlate strongly with \Nhyi\ and \ebv, increasing linearly towards higher density until saturation effect becomes important at \Nhyi$\sim5\times10^{20}\,{\rm cm}^{-2}$ and \ebv$\sim0.08\,$mag. However the observed correlations indicate that the environmental dependence of \nai\ and \caii\ abundance are different. \citet{poznanski12a} performed a similar correlation analysis between \nai\ and \ebv\ and obtained similar results\footnote{For reddening values \ebv\ greater than $0.2$ mag, our results differ from those of \citet{poznanski12a}. We realized that these authors included measurements at high-extinction regions in the Galactic disk due to a few thousand stars erroneously included in the MPA-JHU ``galaxy" catalog. Treating these sources as galaxies contaminates the measurements at high $E(B-V)$ values. These stars have been carefully removed from our analysis.}.
\begin{table*}
\begin{center}
\begin{tabular}{c|c|c|c|c}
\hline
Parameters & Ca K & Ca H & Na D$_2$ & Na D$_1$ \\
\hline
$N_{0}/10^{21}$ ($N_{\rm H\,I}$) & 13.18$\pm$ 4.92  & 32.53$\pm$ 10.83   & 2.08$\pm$ 0.24 & 2.36$\pm$ 0.18    \\
$\alpha_0$ ($N_{\rm H\,I}$) & 0.35$\pm$ 0.03  & 0.39 $\pm$ 0.03  & 0.77 $\pm$ 0.04 & 1.01 $\pm$ 0.04 \\
$E_0$ ($E_{B-V}$) & 4.21 $\pm$ 2.46  & 10.90 $\pm$ 7.47  & 0.39 $\pm$ 0.09   & 0.26 $\pm$ 0.02  \\
$\beta_0$ ($E_{B-V}$) & 0.29 $\pm$ 0.03 & 0.33 $\pm$ 0.04 & 0.63 $\pm$ 0.05  & 1.06 $\pm$ 0.04  \\ \hline
\end{tabular}
\caption{Best-fit parameters for the ${\rm W}_0$-\Nhyi/\ebv\ relations (Equation~\ref{eq:w_nhyi} and \ref{eq:w_ebv}).}
\label{tbl:correlation1}
\end{center}
\end{table*}

We characterize the trends by the following relations:
\beq
W(N_{\rm H\,I}) = \left( \frac {N_{\rm H\,I}}{N_{0}} \right)^{\alpha_0}\;{\rm \AA}  \ \mathrm{,}
\label{eq:w_nhyi}
\eeq
and
\beq
W(E_{B-V}) = \left( \frac{E_{B-V}}{E_{0}}\right) ^{\beta_0}\;{\rm \AA}  \ \mathrm{.}
\label{eq:w_ebv}
\eeq
We estimate the best fit parameters considering only the range where \Nhyi$< 5\times10^{20}\,{\rm cm}^{-2}$ and \ebv$< 0.08$. Above these values the observed trends indicate that the absorption lines become saturated. The best-fit parameters are presented in Table~\ref{tbl:correlation1} and the corresponding trends are overplotted with dash lines in Figure~\ref{fig:CaNa_nh}. Equations \ref{eq:w_nhyi} and \ref{eq:w_ebv} provide us with a way to estimate the \textit{average} absorption strength of \caii\ and \nai\ at a given \Nhyi\ and \ebv. It is interesting to note that the slope of $W_0^{\rm Na\,I}$-\Nhyi\ (and \ebv) relation is close to unity, while the slope of $W_0^{\rm Ca\,II}$-\Nhyi\ is shallower with a best-fit value of about $0.4$. The weaker dependence of the \caii\ absorption on \Nhyi\ could be due to either the lower dust depletion level of calcium or higher ionization parameter or both in lower-density environments. We further investigate the abundances of \caii\ and \nai\ and their dependence on \Nhyi\ in the next section.

\subsubsection{Abundance of \caii\ and \nai}

\begin{figure*}
\includegraphics[width=0.45\textwidth]{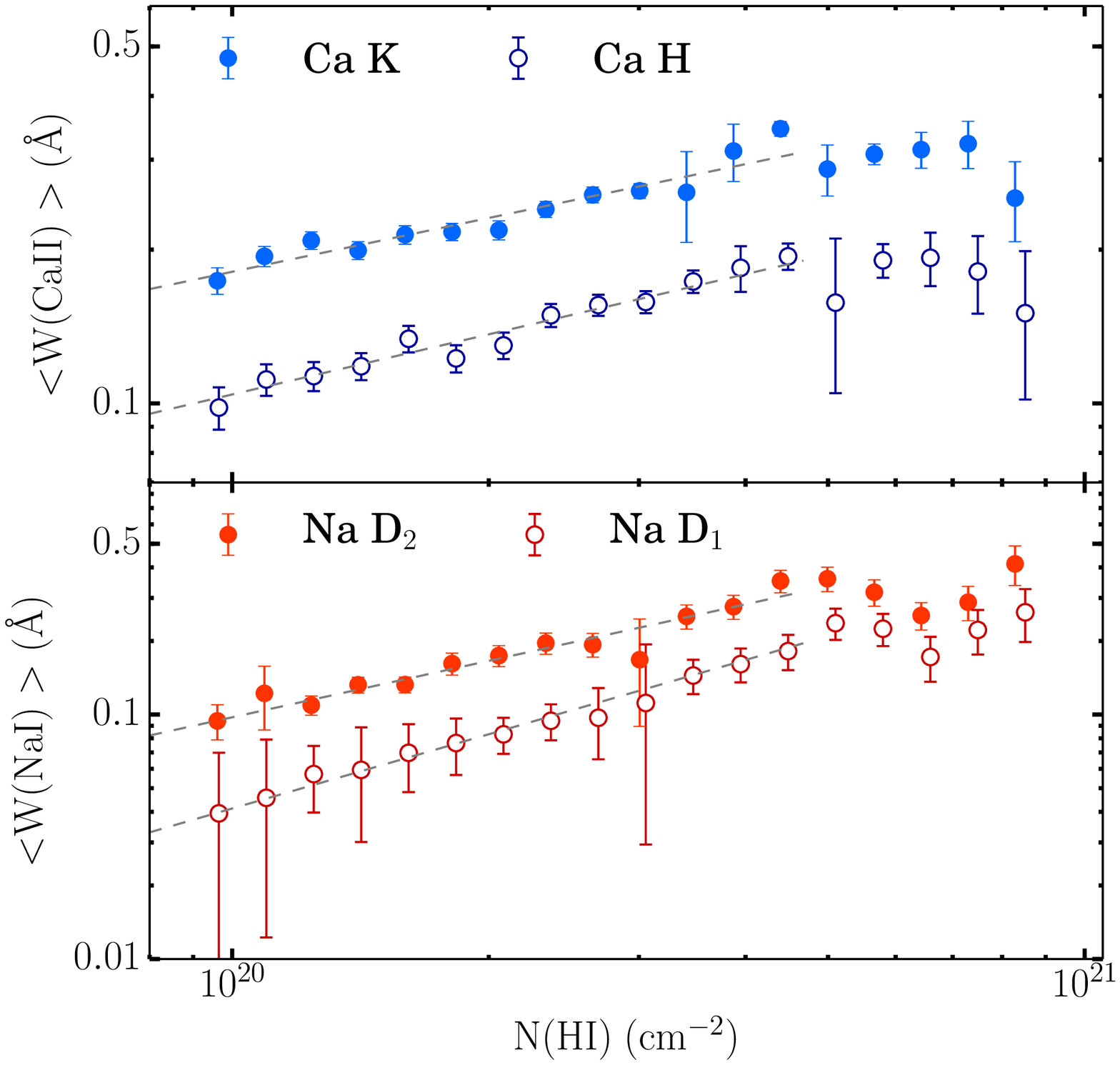}
\includegraphics[width=0.45\textwidth]{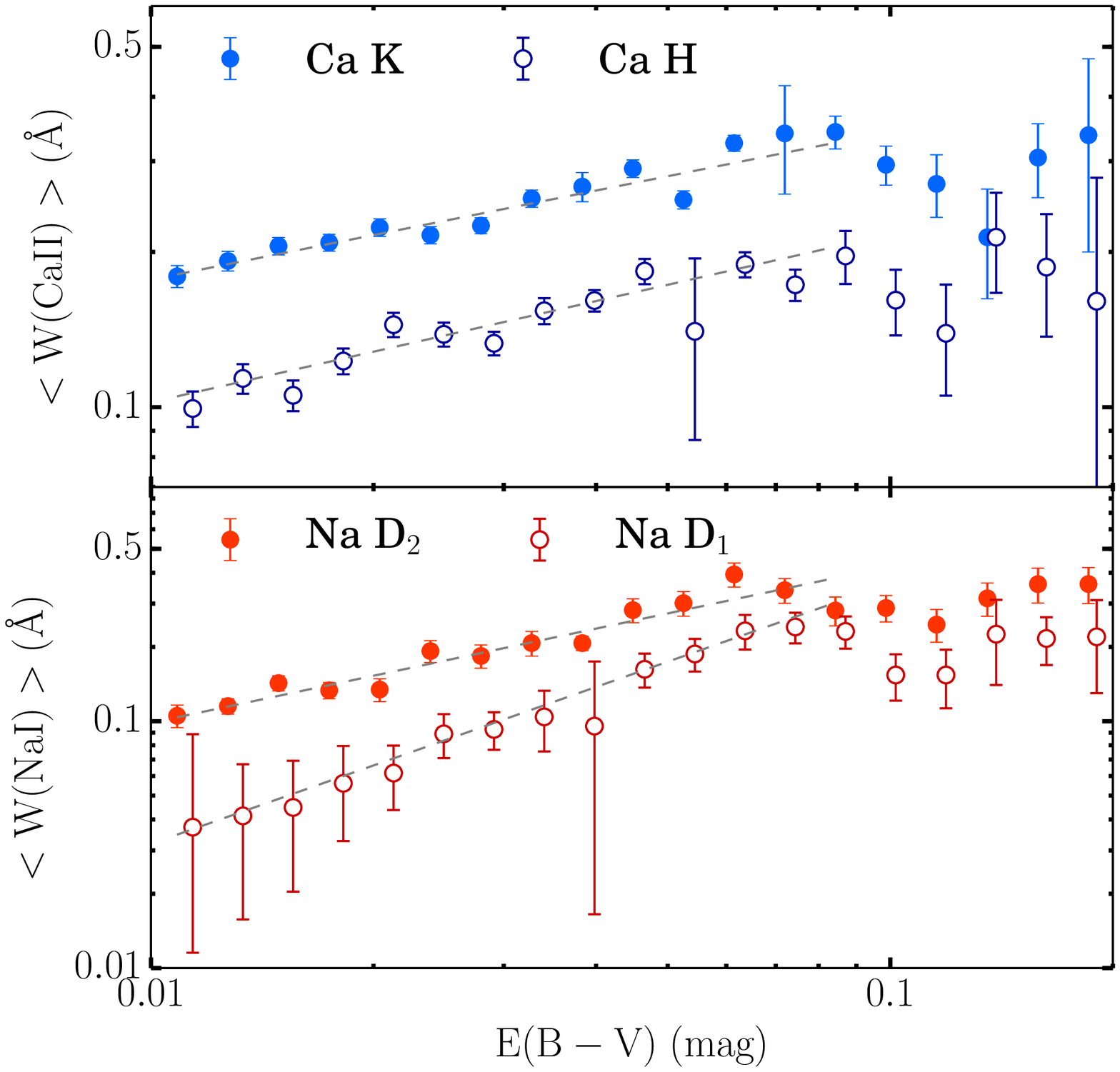}
    \caption{The dependence of \caii\ and \nai\ absorption on neutral hydrogen column density (\Nhyi, left panel) and Galactic reddening (\ebv, right panel). The lines represent the best-fit linear relations with data at $N_{\rm H\,I}<5\times10^{20}\,{\rm cm}^{-2}$ and $E_{B-V}<0.08\,$mag.}
\label{fig:CaNa_nh}
\end{figure*}

\begin{figure*}
    \includegraphics[width=0.4\textwidth]{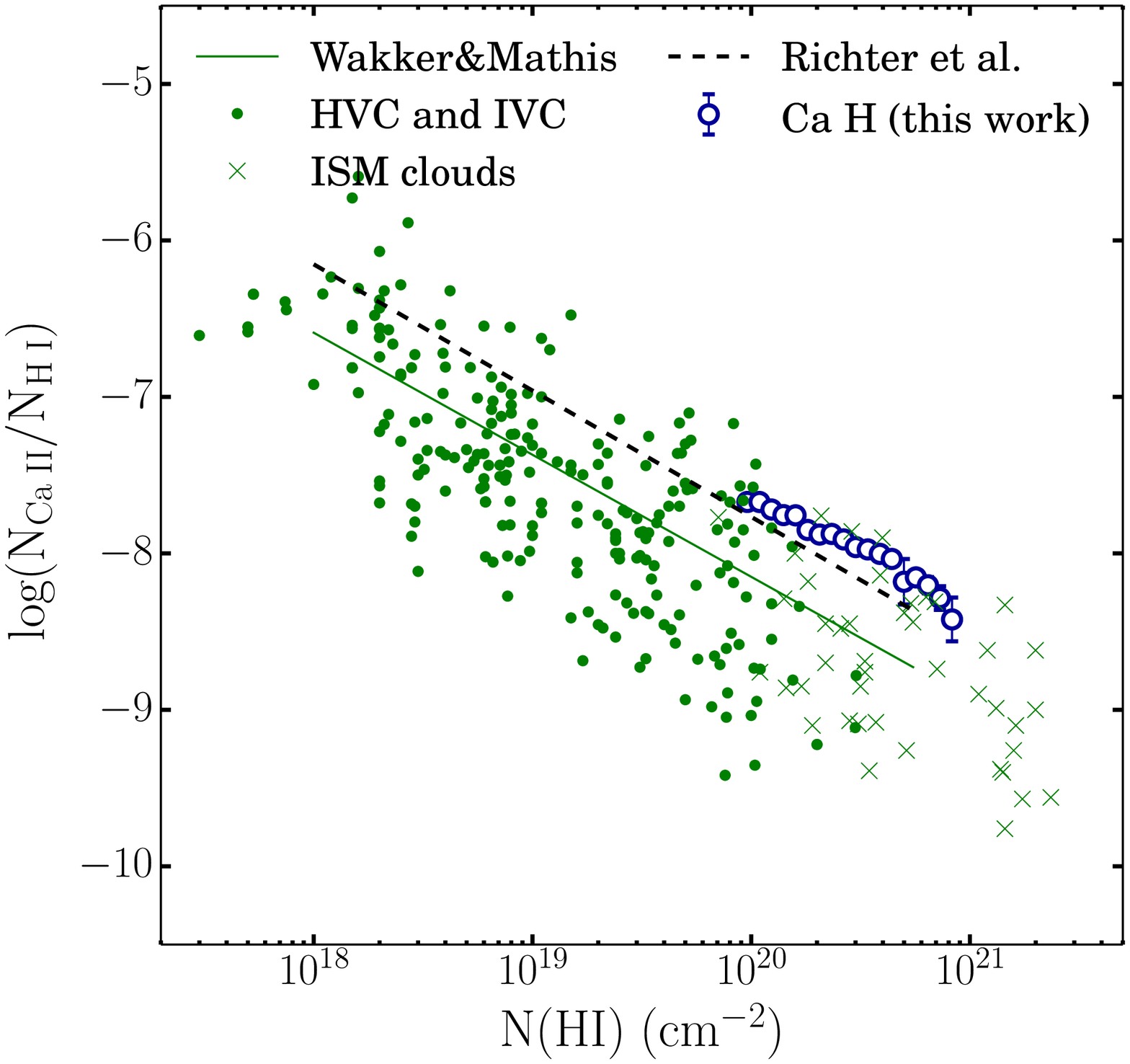}
    \includegraphics[width=0.4\textwidth]{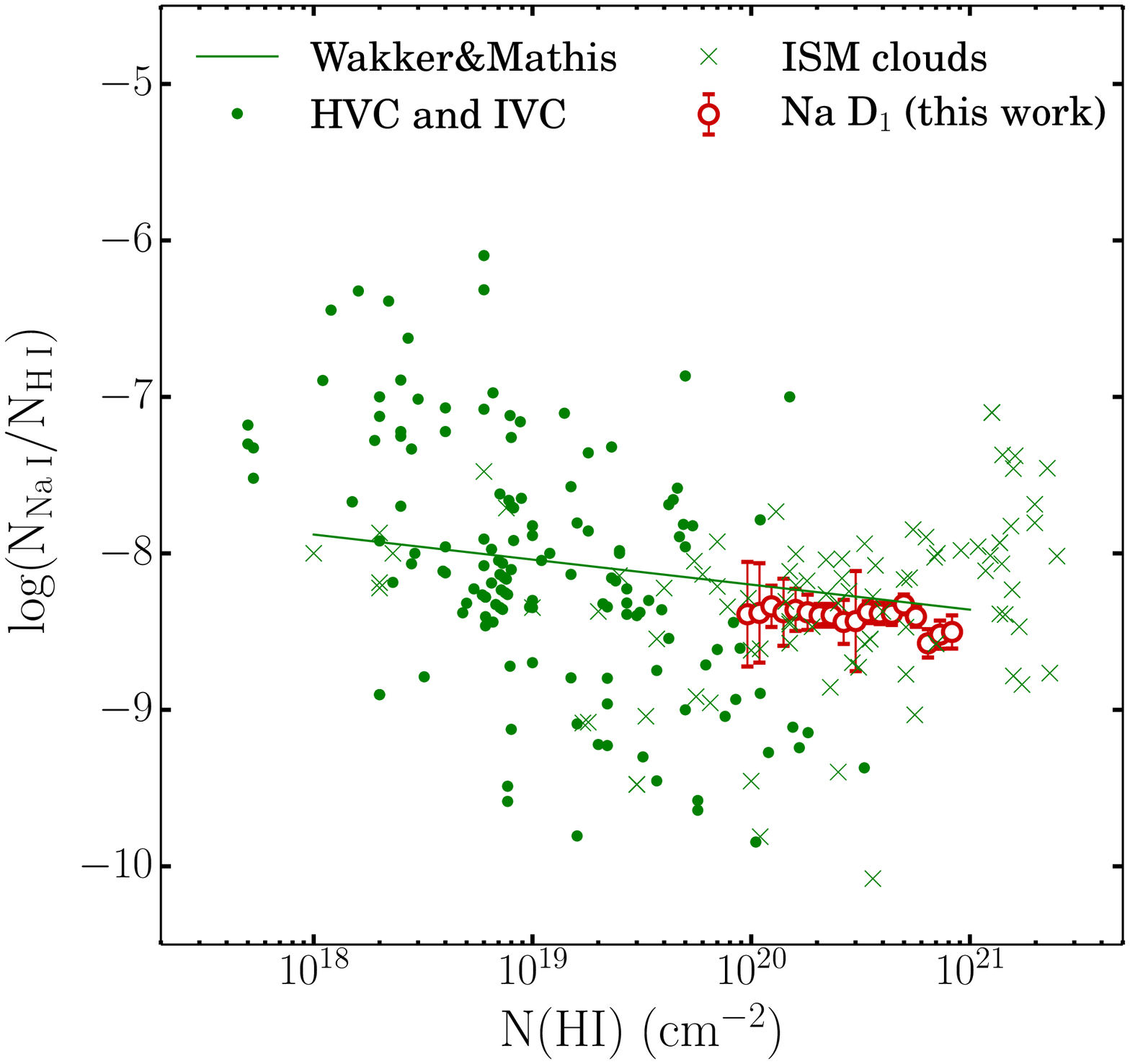}
    \caption{The dependence of \nai\ and \caii\ abundance on neutral hydrogen \Nhyi. The open circles represent the measurements in our \textit{statistical} analysis, integrated over the full velocity range. The dots and crosses show the abundances of \textit{individual} systems, HVC/IVC and ISM clouds, from high-resolution spectroscopy, compiled by \citet{wakker00a}. 
The solid lines are their best-fit linear relations to these individual systems. The dashed line in the left panel shows the best-fit relation to extragalactic \caii\ absorbers by \citet{richter11a}.}
\label{fig:abundance}
\end{figure*}

\begin{figure*}
    \includegraphics[width=0.33\textwidth]{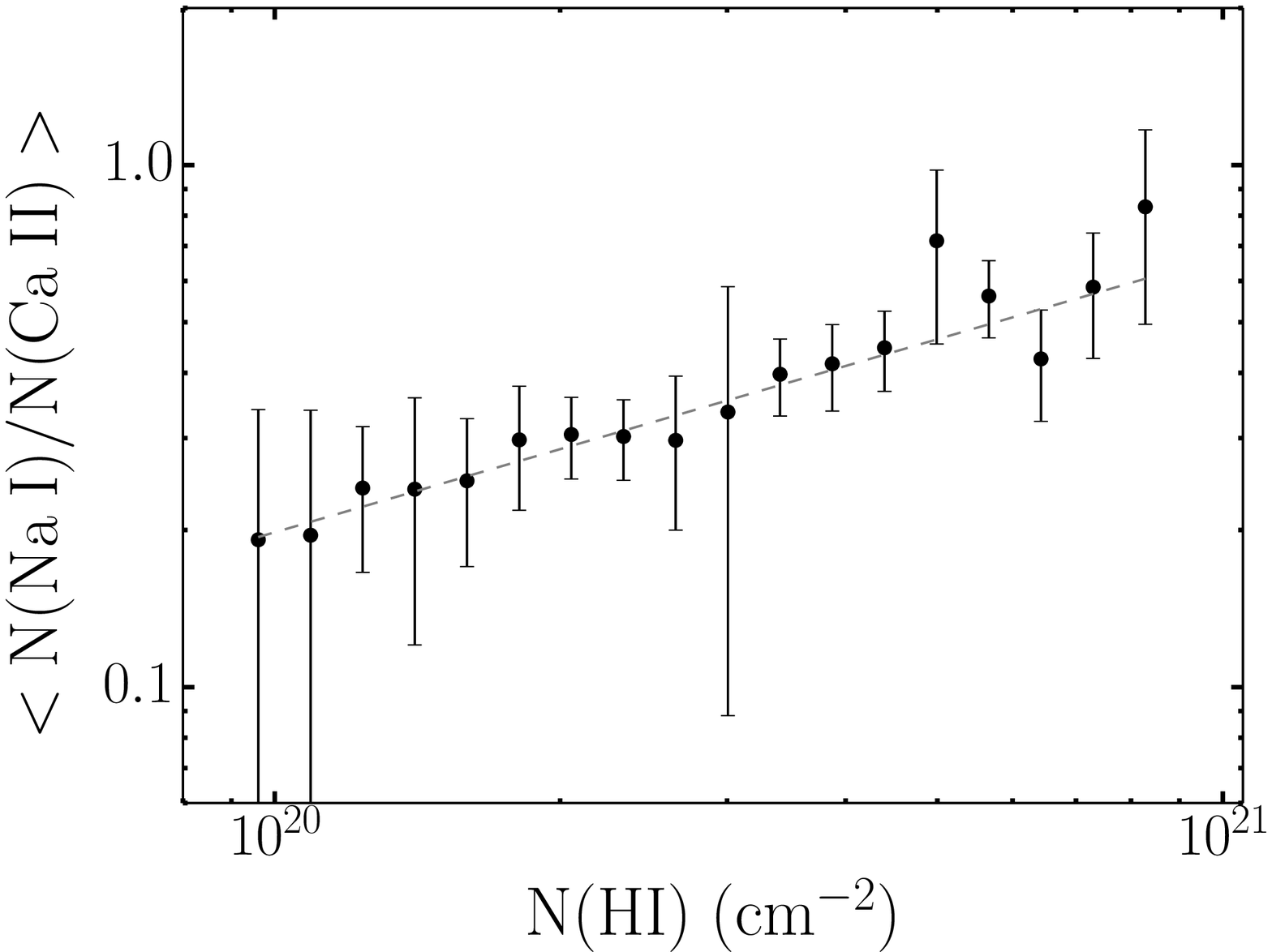}
    \includegraphics[width=0.33\textwidth]{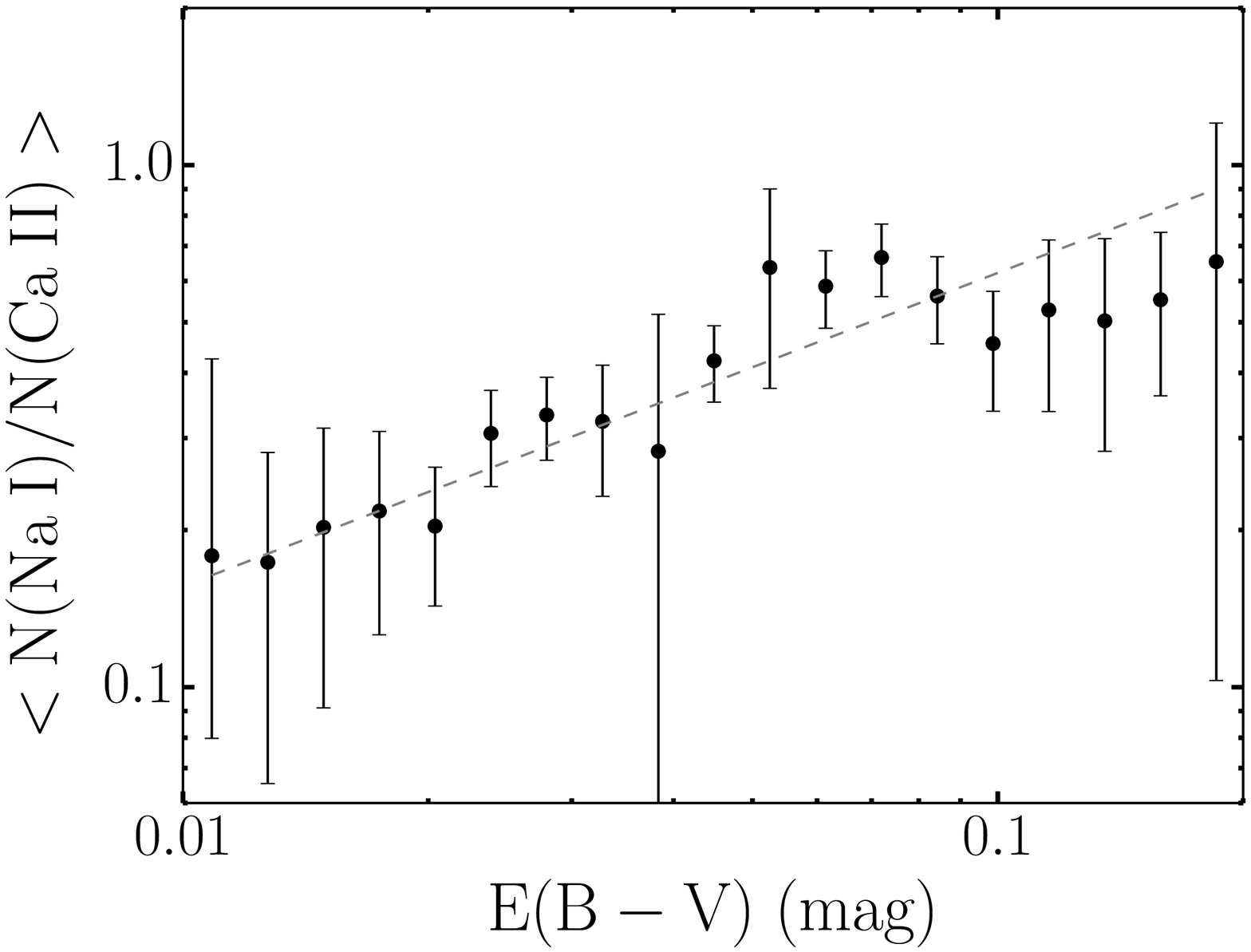}
    \includegraphics[width=0.33\textwidth]{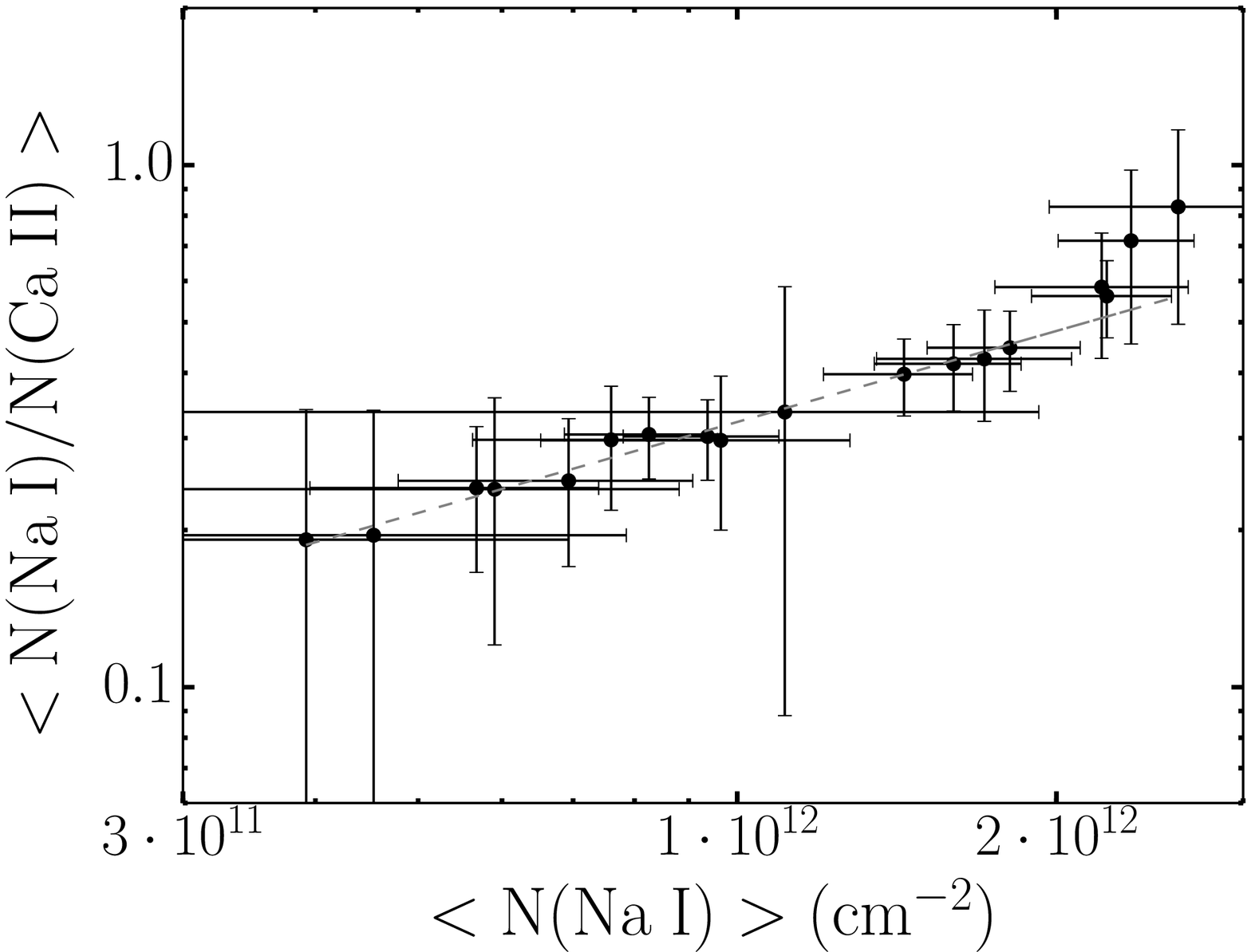}
    \caption{The dependence of the \Nnai/\Ncaii\ ratio on \Nhyi, \ebv\ and \Nnai.}
    \label{fig:ca_na_ratio}
\end{figure*}

We now investigate the dependence of \caii\ and \nai\ abundances on hydrogen column density. We first estimate the \caii\ and \nai\ column density using the \textit{weaker} line of each doublet under the assumption of negligible saturation. Using the linear curve-of-growth relation at low column density, we convert the rest equivalent width of the weaker lines (\rewnaitwo\ and \rewcaiitwo) to column density and calculate the relative abundance $N_{\rm Ca\,II}/N_{\rm H\,I}$ and $N_{\rm Na\,I}/N_{\rm H\,I}$. When the absorption is in the linear regime of the curve of growth, the rest equivalent width is related to the column density with 
\beq
N = 1.13 \times 10^{20}~\mathrm{cm}^{-2}~\frac{W_0}{f\lambda^2}\, \mathrm{,}
\label{eq:density}
\eeq
where both $W_0$ and $\lambda$ are in unit of \AA\ and the oscillator strength $f$ for the \caii\ H line is $0.324$ \citep[][]{safronova11a} and for \nai\ $\mathrm{D}_1$ is $0.320$ \citep[][]{kelleher08a}. This relation is valid when the optical depth is smaller than unity, which corresponds to $N_{\rm Ca\,II, Na\,I} <\sim 10^{12.5}\,{\rm cm}^{-2}$ (assuming the Doppler broadening factor $b\sim10\,\kms$).
The measurements are presented in Figure~\ref{fig:abundance} with open circles. The \nai\ abundance is roughly constant within the \Nhyi\ range probed, while the \caii\ abundance decreases rapidly with increasing \Nhyi.

Our measurements are integrated values over the full velocity range with contributions from both the ISM in the disk and the CGM in the halo. To better understand the physics of the correlations, we compare our measurements with a compilation of abundances of \textit{individual} systems based on high-resolution spectroscopy \citep{wakker00a}. In Figure~\ref{fig:abundance}, the cross symbols represent the measurements of abundance in the ISM clouds, the dotted symbols show the measurements of abundance in high-velocity clouds (HVCs) and intermediate-velocity clouds (IVCs), and the solid lines are the best-fit linear relations by \citet{wakker00a} for these individual systems. In the left panel, we also overplot the best-fit relation by \citet{richter11a} who studied the abundance of \caii\ of about ten \textit{extragalactic} absorber systems. Our measurements of the average \nai\ abundance at a given total \Nhyi\ coincide with the abundance of individual systems, while the average \caii\ abundance is closer to the relation of extragalactic systems and  overall about half a dex higher than those of ISM clouds, though with a similar dependence on \Nhyi.

We would like to stress that the variables in our analysis, the neutral hydrogen column density (\Nhyi), the metal absorption strength and column density, are all integrated values over the full velocity range. The neutral hydrogen column density is dominated by the contribution from the ISM, because along a given line of sight the contribution from the ISM is orders-of-magnitude higher than from the CGM, even if the sightline covers an HVC or IVC. The \nai\ abundance of individual systems, though with a large scatter, on average is roughly constant at different densities. The integrated \nai\ absorption in our analysis therefore must also be dominated by the contribution from the ISM and this explains why the average \nai\ abundance is consistent with that of the individual systems. The \caii\ abundance of individual systems, on the other hand, is orders-of-magnitude higher at low density than at high density. \citet{zhu13b} show that the total amount of \caii\ in the CGM, averaged over all galaxies, is about an order-of-magnitude larger than that in the ISM of the Milky Way \citep[see also][]{richter11a}. In our analysis, making use of extragalactic sources, \caii\ column densities are therefore expected to carry a significant CGM contribution, while hydrogen column densities are mostly originating from the Milky Way disk. We thus expect our measurements to display average values of \caii\ abundance that are higher than those obtained for individual clouds, as shown in figure~\ref{fig:abundance}. 

In Figure~\ref{fig:ca_na_ratio}, we further investigate the dependence of the relative abundance of \nai\ and \caii\ (\Nnai/\Ncaii) on \Nhyi, \ebv\ and \Nnai. The \Nnai/\Ncaii\ ratio is higher at higher hydrogen/dust/\nai\ density, increasing from about $0.4$ at $N_{\rm H\,I}\sim10^{20}\,{\rm cm}^{-2}$ ($E_{B-V}\sim0.01\,{\rm mag}$) to $2$ at $N_{\rm H\,I}\sim10^{21}\,{\rm cm}^{-2}$ ($E_{B-V}\sim0.1\,{\rm mag}$). We fit these relations with similar power-law functions as in Figure~\ref{fig:CaNa_nh}:
\beq
N_{\rm Na\,I}/N_{\rm Ca\,II} = \left( \frac {N_{\rm H\,I}}{N_{1}} \right)^{\alpha_1}\ \mathrm{,}
\label{eq:na_ca_nhyi}
\eeq
and 
\beq
N_{\rm Na\,I}/N_{\rm Ca\,II} = \left( \frac{E_{B-V}}{E_{1}}\right) ^{\beta_1}\ \mathrm{,}
\label{eq:na_ca_ebv}
\eeq
and
\beq
N_{\rm Na\,I}/N_{\rm Ca\,II} = \left( \frac{N_{\rm Na\,I}}{N_{2}}\right) ^{\alpha_2}\ \mathrm{.}
\label{eq:na_ca_nai}
\eeq
The best-fit relations are given in Table~\ref{tbl:correlation2} and overplotted in Figure~\ref{fig:ca_na_ratio}. The slopes are both positive, indicating that the \Nnai/\Ncaii{} ratio is higher in denser environments.

\begin{table}
\begin{center}
\begin{tabular}{c|c}
\hline
Parameters & $N_{{\rm Na\,I}}/N_{{\rm Ca\,II}}$ \\
\hline
$N_{1}/10^{21}$ ($N_{\rm H\,I}$) & 2.13$\pm$ 0.40  \\
$\alpha_1$ ($N_{\rm H\,I}$) & 0.53$\pm$ 0.04  \\
$E_{1}$ ($E_{B-V}$)  & 0.22 $\pm$ 0.06    \\
$\beta_1$ ($E_{B-V}$) & 0.60 $\pm$ 0.07 \\ 
$N_{2}/10^{12}$ ($N_{\rm Na\,I}$) & 7.07$\pm$ 0.82  \\
$\alpha_2$ ($N_{\rm Na\,I}$) & 0.58$\pm$ 0.03  \\ \hline
\end{tabular}
\caption{Best-fit parameters for the \Nnai/\Ncaii\ ratio-\Nhyi/\ebv/\Nnai\ relations (Equation~\ref{eq:na_ca_nhyi}, \ref{eq:na_ca_ebv} and \ref{eq:na_ca_nai}).}
\label{tbl:correlation2}
\end{center}
\end{table}

This environmental dependence of the \Nnai/\Ncaii\ ratio has been noticed for decades \citep[\eg][]{phillips84a, vallerga93a, benbekhti12a}. However, the physical reason is yet to be determined. It is generally attributed to the different dust depletion levels of sodium and calcium at different densities, but other factors such as ionization effects can yield the same dependence. We here briefly discuss two theories. 

A possible scenario was proposed by \citet[][see also \citealt{barlow78c}]{phillips84a}: During the formation of dust grains, sodium is not highly depleted to begin with because of relatively high condensation temperature and low absorbing energy, so density does not play a role in its low depletion level. On the other hand, calcium can be easily accreted onto surface of dust grains due to relatively low condensation temperature and high absorbing energy. At higher density, its accretion rate is higher yet the destruction rate is lower because of greater UV shielding and enhanced mantle growth, while at lower density, calcium trapped in grain surface is easier to be ejected by shock sputtering. All these dynamic processes result in higher calcium depletion level at higher density. 

When an analysis is sensitive to both ISM and CGM absorption,
it is important to take into account of the environmental dependence of metallicity and dust abundance. Gas clouds in the CGM, such as HVCs and the Magellanic Stream, have lower metallicity \citep[about $0.1\,$solar, see][]{wakker99a, richter01a,fox13a} and dust-to-gas ratio \citep[][]{lu98a} than those in the ISM. Although the \textit{total} abundance of Ca and Na is lower in the CGM due to lower metallicity, the overall dust depletion level is also lower because of the low dust content. The combined effects of these dependences therefore can result in the observed dependence of the \Nnai/\Ncaii\ ratio on the environment.

Another theory is based on that the ionization parameter is higher at lower density, and neither \caii\ nor \nai\ is the dominant ionization state. Models with density-dependent ionization parameters but constant dust depletion level can reproduce the dependence of the \Nnai/\Ncaii\ ratio on environment \footnote{Bart Wakker, private communication}. The exact mechanism responsible for this density dependence of \caii\ and \nai\ abundances is still an open question.

\subsection{The Routly-Spitzer effect measured on large scales}

\citet{routly52a} found that the average \Nnai/\Ncaii\ ratio decreases when the cloud velocity
$|v_{\rm LSR}|$ (measured in the local standard of rest) increases. This so-called Routly-Spitzer effect is usually derived from measurements of individual systems in the \textit{local} ISM within tens of kilometer per second \citep[][]{vallerga93a}. Recently, \citet{benbekhti12a} 
reported a similar effect
for individual absorption systems, based on deviation velocity measurements\footnote{Deviation velocity is the difference of the radial velocity of the cloud and the terminal velocity of the Milky Way disc \citep[][]{wakker91a}. However, for high galactic latitude systems, the effect should be similar regardless of the rest frame.} in the halo of the Milky Way, extending the detection of the Routly-Spitzer effect to larger scales. Here we investigate if such an effect also exists in our statistical data.

\begin{figure}
    \includegraphics[width=0.4\textwidth]{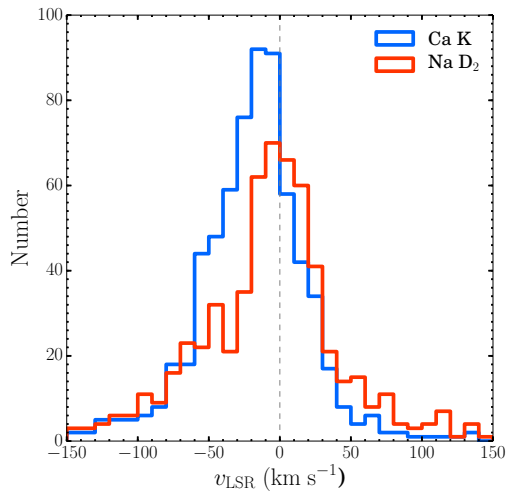}
    \caption{The velocity distributions of \caii\ and \nai\ absorption in the local standard of rest using measurements from the northern Galactic hemisphere. The distribution for \caii\ is offset from the origin and wider than that of \nai. This is consistent with the interpretation that \nai\ originates mostly from the disk while \caii\ also traces a significant amount of material from the circumgalactic medium.
}
\label{fig:vlsr_stats}
\end{figure}

To study the dependence of the \Nnai/\Ncaii\ ratio on \vlsr, we use the HEALPix maps as presented in Figure~\ref{fig:maps}. In each HEALPix pixel, we determine the \emph{centroid} wavelength from the double-Gaussian fitting and calculate the velocity offset from the rest-frame wavelength of the doublets. When doing so we find that \nai\ and \caii\ are on average blueshifted by $8$ and $23\,\kms$ in the heliocentric reference frame. We then convert these observed velocities to the local standard of rest \citep[LSR, ][]{corteau}. We show the corresponding distributions of absorption velocities for \caii\ and \nai\ in the northern hemisphere in Figure~\ref{fig:vlsr_stats}. As can be seen, \caii\ tends to be found at lower velocities compared to \nai. We find the mean of the \nai\ velocity distribution to be about $-3\,\kms$, while the value for \caii\ is found to be about $-17\,\kms$. 
Recent \hyi\ surveys have shown that, in the northern sky, the high- and intermediate-velocity clouds have on average negative velocities \citep[][]{richter03a, putman12a}. The higher detection rate of \caii\ at 
negative velocities is therefore consistent with the interpretation that \caii\ detected in the extragalactic sources traces a significant amount of gas from the CGM.

\begin{figure*}
    \includegraphics[width=0.3\textwidth]{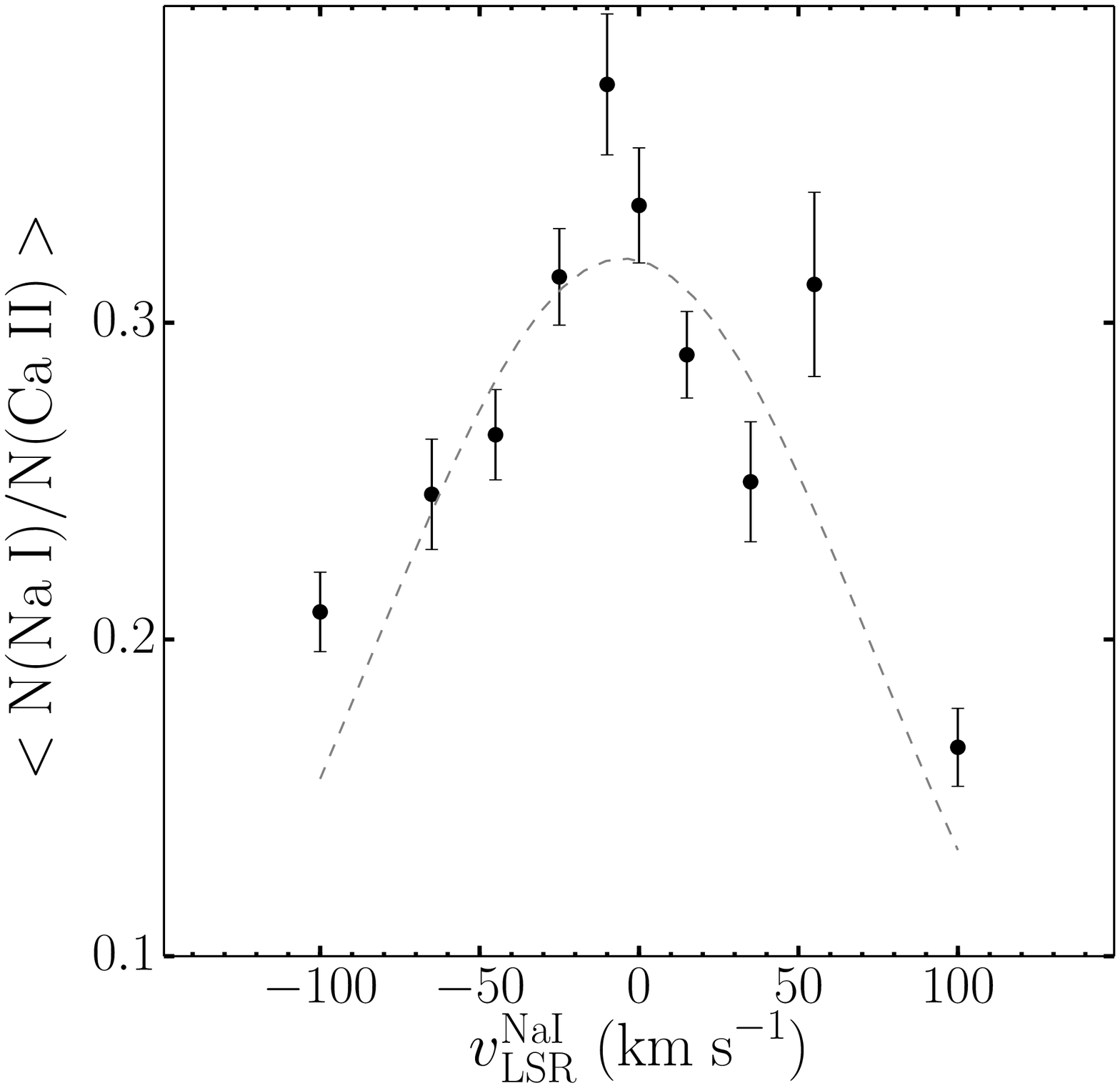}
    \includegraphics[width=0.3\textwidth]{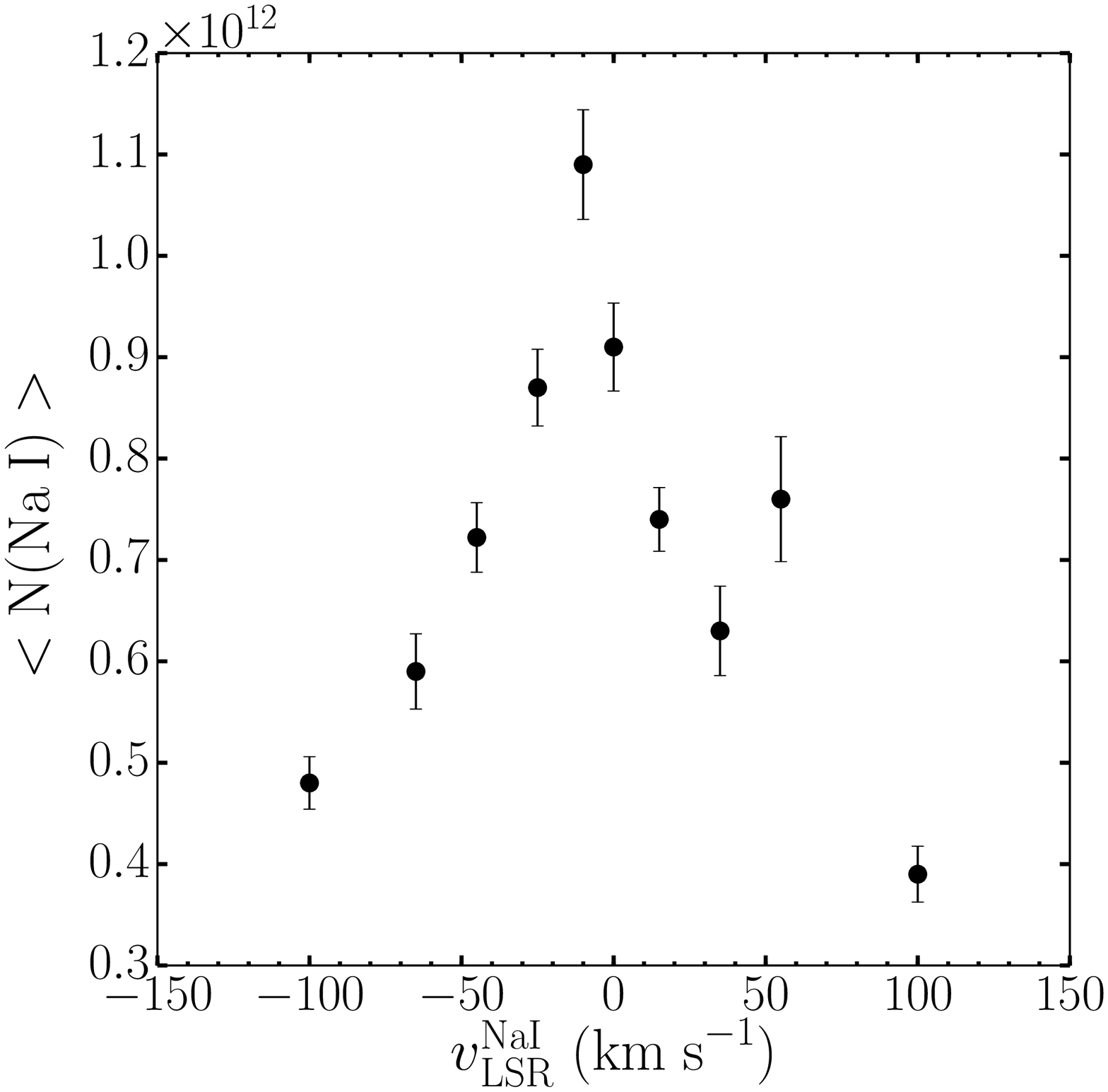}
    \includegraphics[width=0.3\textwidth]{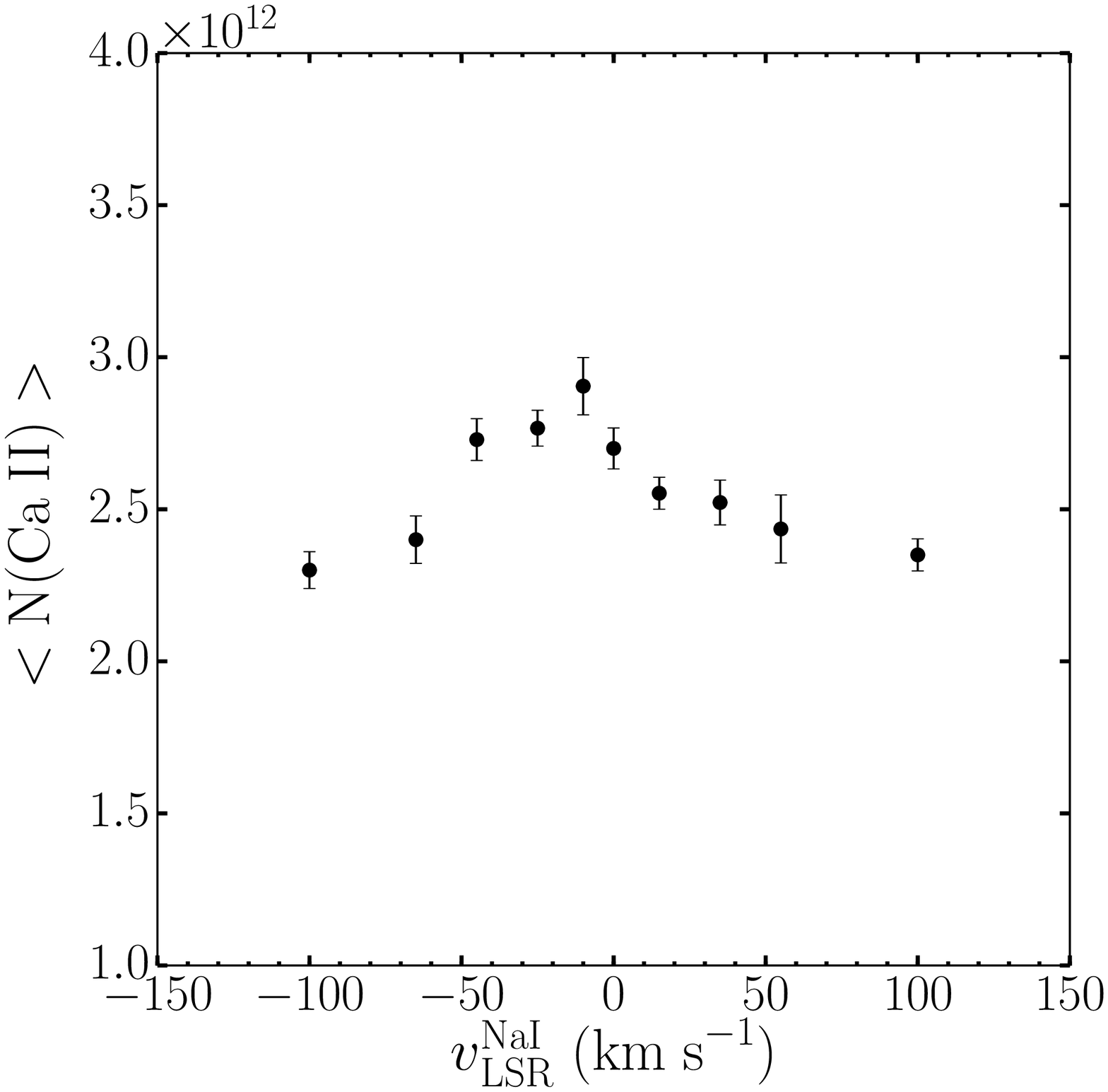}
    \caption{The global Routly-Spitzer effect. The left panel shows the dependence of the \Nnai/\Ncaii\ ratio on \vlsr\ of \nai. The right panels show \Nnai\ and \Ncaii\ as a function of \vlsr. Note the y-axis ranges in the right two panels are consistent, i.e., the maximum is a factor of four of the minimum in both cases.}
    \label{fig:naca_vlsr}
\end{figure*}

We then group the pixels with similar $v_{\rm LSR}^{\rm Na\,I}$ values and measure the mean $W_0^{\rm Na\,I}$ and $W_0^{\rm Ca\,II}$ in each $v_{\rm LSR}^{\rm Na\,I}$ bin. We note that using the velocity derived from \caii\ produces similar trends with a larger scatter. The left panel of Figure~\ref{fig:naca_vlsr} shows our measurements of the \Nnai/\Ncaii\ ratio as a function of $v_{\rm LSR}^{\rm Na\,I}$. Interestingly, we find that the average \Nnai/\Ncaii\ ratio shows a similar dependence on \vlsr\ as the \textit{local} Routly-Spitzer effect for individual systems, therefore our data show that the velocity dependence of \Nnai/\Ncaii\ ratio also exists on the Galactic scale.
\citet{barlow77a} suggested that it is related to dust destruction. One of the major mechanisms of dust destruction is sputtering by interstellar shocks. The destruction rate depends on the shock velocity which results in higher dust destruction rate and higher \caii\ abundance relative to \nai\ at higher velocity. Subsequent theoretical analyses and simulations have lent support to such proposition \citep[\eg][]{draine79a, mckee89a, jones94a}. However, it is also possible that this effect can be just a consequence of the density dependence of the \Nnai/\Ncaii\ ratio shown in the previous section, regardless of the actual cause. In the middle and right panels of Figure~\ref{fig:naca_vlsr}, we show that the \nai\ density has a similar dependence on \vlsr\ as the \Nnai/\Ncaii\ ratio, while \caii\ density has a much weaker dependence. Since the \Nnai/\Ncaii\ ratio is higher in higher-density environment, as probed by \nai, \hyi\ and dust (Figure~\ref{fig:ca_na_ratio}), this can therefore explain part of the velocity dependence. The mechanisms responsible for the Routly-Spitzer effect on large scales can therefore be more complicated than interstellar shock only.

\section{Implications for photometric observations and redshift estimation}
\label{sec:photometry}

The \caiidoublet\ and \naidoublet\ absorption lines due to the ISM and CGM of the Milky Way are expected to impact the broad band photometry of extragalactic objects. Two types of effects can occur:
\begin{itemize}
\item absorption on top of continuum flux: Metal line absorption is expected to suppress a fraction of the expected flux from extragalactic sources.
This effect has been statistically detected in the broad band photometry of distant quasars \cite{menard2012}.
We note that the distribution of metals in the gas phase does not \textit{exactly} track that of the dust. Therefore infrared emission (as used in SFD) cannot be used to precisely map this kind of absorption.
\item Absorption at the location of emission lines:
if a redshifted emission line happens to coincide with one of the Galactic absorption features, the apparent broad-band color of the corresponding galaxy can be appreciably modified. This can happen when, for example,  \oiiilam\ is seen on top of Na D at $z\sim0.176$ or \oiilam\ is seen on top of Na D at $z\sim0.58$.
This introduces a level of degeneracy between redshift and color for star forming galaxies and is expected to impact the accuracy of photometric redshift estimation (see Rahman et al. (in prep) for a more extended discussion of this effect).
\end{itemize}

\section{Summary}\label{sec:summary}

The light emitted by extragalactic sources has to penetrate the ubiquitous gas and dust in the interstellar and circum-galactic media of the Milky Way. Every extragalactic spectrum therefore carries information about these two environments. 

We have used the large collection of quasar and galaxy spectra obtained by the Sloan Digital Sky Survey to detect absorption lines induced by the presence of \caii\ and \nai. The absorption signal is typically too weak to be detected in individual SDSS spectra. Measuring it requires a statistical approach. Not until recently had this avenue been explored \citep[\eg][]{poznanski12a,2014arXiv1406.7284L,2015ApJ...798...35Z, 2015MNRAS.447..545B}. 
We have created maps revealing the spatial distribution of \caii\ and \nai\ absorption on the sky. We have then used these new datasets to study correlations between these two components and other baryonic tracers. 
Our analysis has allowed us to obtain the following results:
\begin{itemize}
\item We have mapped out the \nai\ D and \caii\ H \& K absorption over about a quarter of the sky, obtaining signal-dominated measurements at a resolution of about one degree. The corresponding maps are publicly available. 
\item Cross-correlating these absorption maps with \hyi\ column density and dust reddening maps, we
find that the dependence of \caii\ absorption on \Nhyi\ and \ebv\ is much weaker than \nai and provide fitting formulae for the average relations between these quantities. We find the abundance of \nai\ with respect to neutral hydrogen to be roughly constant in different environments, while the \caii\ abundance decreases with hydrogen column density. We discuss the possible mechanisms that can cause such different behaviors, including dust depletion and ionization effects.
\item We show that, on average, the \Nnai/\Ncaii\ ratio decreases with velocity with respect to the local standard of rest up to velocities reaching $100\,\kms$. Interestingly, this is consistent with the local Routly-Spitzer effect originally measured for individual systems within a much smaller velocity range.
\item We commented on the fact that Galactic absorption lines can change the apparent colors of emission line galaxies at specific redshifts, which can possibly impact photometric redshift estimation.
\end{itemize}

The detection of absorption features relies on the ability to accurately estimate the intrinsic spectral energy distribution of the background source. To do so we have used the continuum estimates of quasars presented in \citet{zhu13a}, based on a vector-decomposition method, and of galaxies by \citet{zhu10a}, based on stellar spectral templates. Our work demonstrates that large spectroscopic surveys, though designed for studies of the light sources themselves or cosmology, provide valuable datasets for studies of the ISM and CGM of our own Milky Way. We have only just started to explore this avenue and there is more information to be extracted. For example, the absorption induced in the spectra of extragalactic sources includes contributions from both the ISM and CGM of the Milky Way, while absorption in the stellar spectra only has contribution from the ISM up to the star's distance, differential analysis therefore can be performed to produce a 3D map of the gas distribution in and around the Milky Way. 
The continuum-estimation methods and statistical absorption-line spectroscopic approaches we developed are robust, generic, and readily applicable to any large datasets from ongoing and future surveys, such as SDSS-IV, DESi and PFS.

\vspace{0.15in}
\section*{Acknowledgments}

We thank Philipp Richter and Bart Wakker for useful discussions. This work was supported by NSF Grant AST-1109665, the Alfred P. Sloan foundation and a grant from Theodore Dunham, Jr., Grant of Fund for Astrophysical Research and RFBR grant 14-02-31456. G.Z. acknowledges partial support provided by NASA through Hubble Fellowship grant \# HST-HF2-51351 awarded by the Space Telescope Science Institute, which is operated by the Association of Universities for Research in Astronomy, Inc., under contract NAS 5-26555.

Funding for the SDSS and SDSS-II has been provided by the Alfred P. Sloan Foundation, the Participating Institutions, the National Science Foundation, the U.S. Department of Energy, the National Aeronautics and Space Administration, the Japanese Monbukagakusho, the Max Planck Society, and the Higher Education Funding Council for England. The SDSS Web Site is http://www.sdss.org/. Funding for SDSS-III has been provided by the Alfred P. Sloan Foundation, the Participating Institutions, the National Science Foundation, and the U.S. Department of Energy Office of Science. The SDSS-III web site is http://www.sdss3.org/.

\bibliographystyle{mn2e.bst}

\end{document}